\documentclass[conference]{IEEEtran}
\IEEEoverridecommandlockouts
\usepackage{cite}
\usepackage{amsmath,amssymb,amsfonts}
\usepackage{algorithmic}
\usepackage{graphicx}
\usepackage{textcomp}
\usepackage{pifont}
\usepackage{xcolor}
\usepackage{wasysym}
\usepackage{flushend} 
\usepackage{multirow}
\usepackage[normalem]{ulem}
\useunder{\uline}{\ul}{}
\usepackage{acronym}
\usepackage[citecolor=black,
            colorlinks=true,
            linkcolor=black,
            urlcolor  = black,
            pdftitle={Face-Voice Association for Audiovisual Active Speaker Detection in Egocentric Recordings},
            pdfauthor={Jason Clarke, Yoshihiko Gotoh, and Stefan Goetze},
            pdfkeywords={Diarization, Audiovisual Active Speaker Detection, Video-based Face Recognition, Speaker Recognition},
            pdfsubject={Diarization, Audiovisual Active Speaker Detection, Face-voice Association}]{hyperref}
\usepackage[english]{babel}
\addto\extrasenglish{}
\addto\extrasenglish{}
\addto\extrasenglish{}
\addto\extrasenglish{}
\addto\extrasenglish{}
\addto\extrasenglish{}
\acrodef{AR}    {augmented reality}
\acrodef{ASD}   {audiovisual active speaker detection}
\acrodef{AVD}   {audiovisual diarization}
\acrodef{GRU}   {gated recurrent unit}
\acrodef{CP}    {conversation Participant}
\acrodef{DeiT}  {data-efficient image transformer}
\acrodef{FoV}   {field of view}
\acrodef{FLOPs} {floating point operations}
\acrodef{mAP}   {mean Average Precision}
\acrodef{MFCC}  {Mel frequency cepstral coefficient}
\acrodef{MLP}   {multilayer perceptron}
\acrodef{SCAN}   {speaker comparison auxiliary network} 
\acrodef{SNR}   {signal-to-noise-ratio}
\acrodef{SoTA}  {state-of-the-art}
\acrodef{SL-ASD} {Self-Lifting for audiovisual active speaker detection}
\acrodef{VAD}   {voice activity detector}
\acrodef{V-TCN} {visual temporal convolutional network}

\def\BibTeX{{\rm B\kern-.05em{\sc i\kern-.025em b}\kern-.08em
    T\kern-.1667em\lower.7ex\hbox{E}\kern-.125emX}}
\begin{document}

\title{Face-Voice Association for Audiovisual Active Speaker Detection in Egocentric Recordings\\

\thanks{This work was supported by the Centre for Doctoral Training in Speech and Language Technologies (SLT) and their Applications funded by UKRI [grant number EP/S023062/1]. This work was also funded in part by Meta.}
}

\author{
  \IEEEauthorblockN{
    Jason Clarke$^1$, 
    Yoshihiko Gotoh$^1$, and 
    Stefan Goetze$^{1,2}$
  }
  \IEEEauthorblockA{
    $^1$Speech and Hearing (SPandH) group, School of Computer Science,
    The University of Sheffield, Sheffield, United Kingdom
  }
  \IEEEauthorblockA{
    $^2$South Westphalia University of Applied Sciences, Iserlohn, Germany
  }
  \IEEEauthorblockA{
    \{jclarke8, y.gotoh, s.goetze\}@sheffield.ac.uk, goetze.stefan@fh-swf.de
  }
}
\maketitle

\begin{abstract}
\Ac{ASD} is conventionally performed by modelling the temporal synchronisation of acoustic and visual speech cues. In egocentric recordings, however, the efficacy of synchronisation-based methods is compromised by occlusions, motion blur, and adverse acoustic conditions. In this work, a novel framework is proposed that exclusively leverages cross-modal face-voice associations to determine speaker activity. An existing face-voice association model is integrated with a transformer-based encoder that aggregates facial identity information by dynamically weighting each frame based on its visual quality. This system is then coupled with a front-end utterance segmentation method, producing a complete \ac{ASD} system. This work demonstrates that the proposed system, \ac{SL-ASD}, achieves performance comparable to, and in certain cases exceeding, that of parameter-intensive synchronisation-based approaches with significantly fewer learnable parameters, thereby validating the feasibility of substituting strict audiovisual synchronisation modelling with flexible biometric associations in challenging egocentric scenarios.
Code is available at \linebreak\url{https://github.com/jclarke98/SL_ASD}.
\end{abstract}


\vspace{-0.5em} 

\section{Introduction}
\Acf{ASD} refers to the task of identifying the video-framewise speaking activity of a candidate speaker through the joint analysis of an audio signal and its temporally aligned face track\cite{activespeakersincontext, ava-as, asdtransformer, buffy, talknet, Liao_2023_CVPR,ASDNet}. While a candidate speaker is active, speech is present in the audio signal and accompanied by visual speech-related cues, such as lip movement or cheek posture \cite{asdtransformer}, in the video signal. These visual cues are observed in the face-track, which is defined as a temporally contiguous set of face crop frames focused on a single candidate speaker. Crucially, these audiovisual cues must exhibit temporal alignment if the candidate speaker is the source of the speech.

Modern approaches to \ac{ASD} predominantly follow the single-candidate paradigm which relies on this alignment: modelling the synchronisation between speech present in the audio signal and the visual cues indicative of speech present in the video signal\cite{talknet, sync-talknet, asdtransformer, Liao_2023_CVPR, buffy, lookwhostalking}. Recent advancements extend this single-candidate paradigm by incorporating additional contextual information like inter-speaker relationships\cite{activespeakersincontext, EASEE, SPELL, ASDNet, LeonAlcazar2021MAASMA}, full-scene image modelling\cite{clarke23-ASD-ASRU}, and speaker-specific information\cite{TS-talknet, clarke2025speakerembeddinginformedaudiovisual}. These advancements address challenges posed by multi-talker scenarios and noisy environments. While such extensions have yielded incremental performance improvements, they remain fundamentally dependent on audiovisual cues, indicative of speech, being temporally correspondent.

In egocentric recording scenarios, it has been shown that the efficacy of synchronisation-based approaches is significantly diminished. This is because fine-grained visual speech cues required for synchronisation-based approaches are often compromised by occlusions, motion blur, or adverse lighting conditions. Additionally, the audio signals are susceptible to interference from overlapping speech or environmental noise. These factors degrade the reliability of both modalities, contributing to a pronounced performance disparity when comparing \ac{ASD} systems on egocentric\cite{Ego4D, spellego4dchallenge} vs.\ exocentric benchmarks\cite{ava-as}.
 
An alternative approach emerges from face-voice association methods, which link facial identities to voice profiles without requiring strict audiovisual synchronisation\cite{seeking_shape_of_sound, Self-Lifting, learnable_pins, single-branch, disentangled_cross-modal_biometric_matching, momentum_face-voice}. Unlike traditional \ac{ASD} approaches, which require a majority of crisp frames, these methods operate frame-wise for facial analysis, requiring only a single high-quality frame with resolvable identity cues to establish a cross-modal association. This characteristic proves advantageous in egocentric settings, where transient moments of clear facial visibility can suffice for robust associations. Prior work has explored related ideas: for instance, permutation-invariant unimodal similarity matrices between face and speaker embeddings have been proposed for utterance attribution\cite{unimodal_similarity}, though these lack joint cross-modal training to learn a unified biometric space. Others have integrated face-voice association as an auxiliary component to synchronisation-based \ac{ASD} systems\cite{FaVoA}, but such hybrid approaches underperform their baseline models even in exocentric contexts~\cite{ava-as}.

This work diverges from previous methods~\cite{FaVoA, unimodal_similarity} by exclusively leveraging face-voice association as the foundation for \ac{ASD}, circumventing the  reliance on synchronisation-based modelling altogether. The presented system adapts an existing face-voice association model and integrates it with a front-end utterance segmentation system. This integrated system, namely \acf{SL-ASD}, enables the attribution of utterance-level speech segments to their corresponding visually identifiable speakers. Critically, this work incorporates a transformer encoder that aggregates the temporal context provided by video data, dynamically weighting each frame’s contribution based on its visual quality to produce a single discriminative identity embedding~\cite{clarke2025speakerembeddinginformedaudiovisual}. This aggregation leverages video data without reintroducing dependence on fine-grained visual cues—a vulnerability of synchronisation-based methods.

The approach presented in this work demonstrates that purely cross-modal identity associations, when enhanced with frame-quality informed aggregation, and a robust segmentation front-end achieves performance comparable to parameter-intensive synchronisation-based methods in egocentric settings. This work underscores the viability of withdrawing from approaches primarily dependent on audiovisual temporal correspondence modelling in the challenging context of egocentric recordings.

\section{Face-Voice Association For Audiovisual Active Speaker Detection}
\label{sec:method}
In the following, synchronisation-based \ac{ASD} is replaced by \ac{SL-ASD}, which consists of a two-stage framework: (1) speaker-invariant utterance segmentation and (2) face-voice association with frame-quality informed aggregation. Unlike conventional \ac{ASD} systems that model temporal audiovisual correspondence, the \ac{SL-ASD} approach exploits cross-modal identity associations. This design directly addresses egocentric challenges where transient visual clarity suffices for biometric linking, but not for synchronisation-based approaches. 


\begin{figure}[!htb]
  \centering
    \includegraphics[scale=0.08]{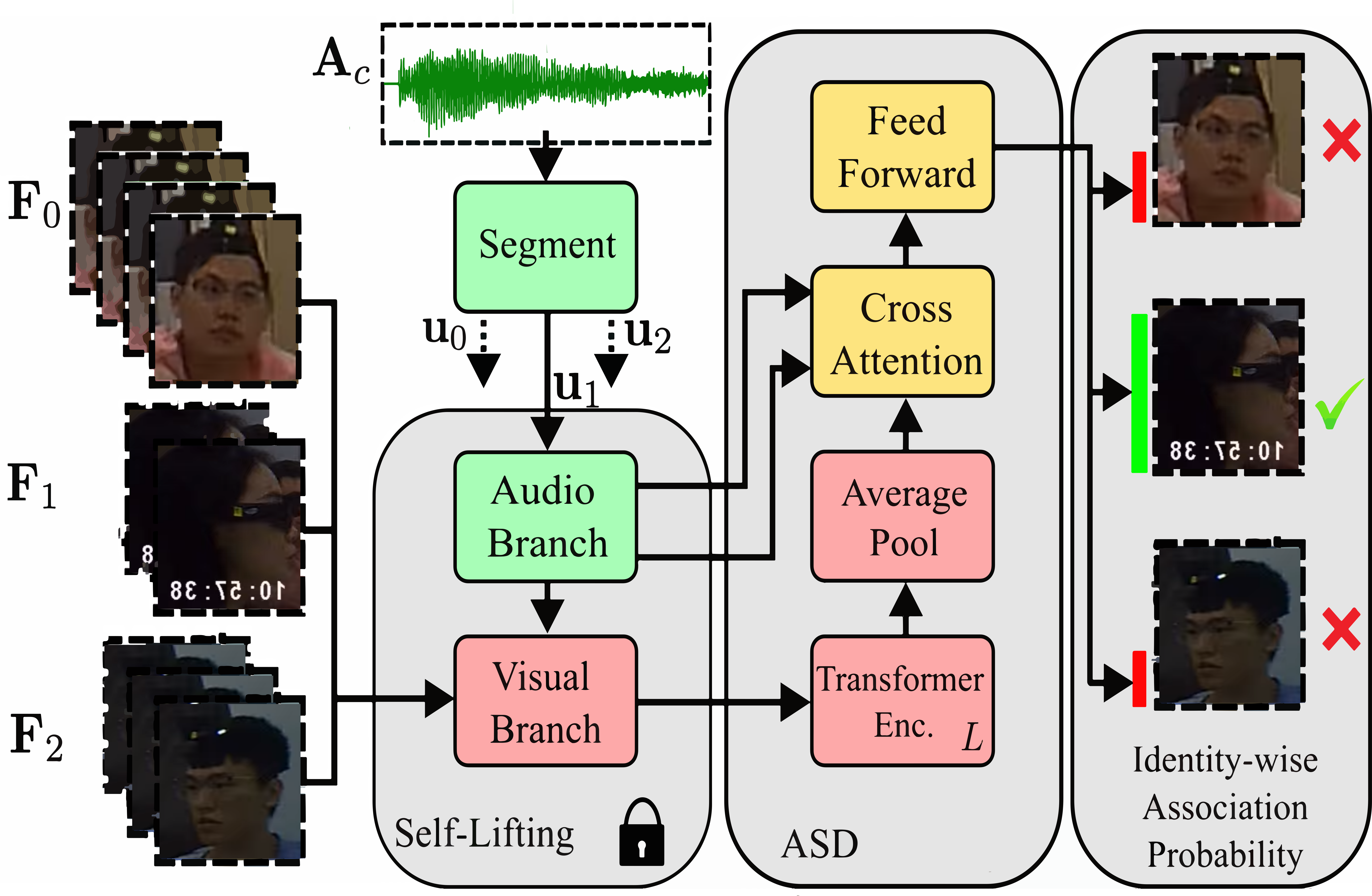}
    \caption{\ac{SL-ASD} framework, dotted lines of utteraces indicate only utterances belonging to a single speaker are passed through the pipeline at a time during training, in this case $\mathbf{u}_1$. Colours indicate modality. Bars adjacent to end faces indicate probability of a face-voice match.}
\label{fig:SL_ASD}
\end{figure}

\vspace{-0.5em} 
\subsection{Notation and Overview} 
Typically, egocentric datasets, such as Ego4D\cite{Ego4D}, consist of a collection of video clips denoted by $\mathcal{C}$. Each video clip $c \in \mathcal{C}$ comprises two primary components: an audio signal $\mathbf{A}_c$ and a video signal $\mathbf{V}_c$. The video component is further decomposed into a set of face tracks, denoted by $\mathcal{F}_c$. Each face track corresponds to a sequence of temporally contiguous face crop bounding boxes that focus on a single identity and is represented by the tensor $\mathbf{F}_{i,j} \in \mathbb{R}^{T_{i,j} \times C \times H \times W}$, where $i$ is the identity index, $j$ is track index, $T_{i,j}$ is the number of frames in the face track, $C$ is the number of channels, and $H$ and $W$ denote the height and width of each bounding box, respectively.

Similarly, the audio component of each video clip is decomposed into a set of utterances, denoted by $\mathcal{U}_c$. An utterance is defined as a short segment of speech spoken by a single speaker and is represented by the vector $\mathbf{u}_{i,k} \in \mathbb{R}^{T_{A_k}}$, where $k$ is the index for the utterance and $T_{A_k}$ denotes its duration in audio samples.

\vspace{-0.5em} 
\subsection{Speaker-Invariant Utterance Segmentation}
The correspondence between an individual's visual appearance and vocal characteristics has been well established in the literature\cite{what_do_we_hear, infant_fva}. Face-voice association frameworks have been developed to exploit this relationship by learning joint embedding spaces in which speaker embeddings and face embeddings corresponding to the same identity are positioned in close proximity, while embeddings of different identities are separated by greater distances\cite{disentangled_cross-modal_biometric_matching, Self-Lifting, learnable_pins}. This principle is leveraged in this work and applied to the task of \ac{ASD}.

Speaker-invariant utterances are first extracted from the audio component of each video clip $\forall \mathbf{A}_c$. The performance of the front-end utterance segmenter is critical to the overall effectiveness of the framework. Errors such as missed detections, false detections, and utterances containing overlapping speech propagate uncorrected throughout the pipeline, leading to significant degradation in system performance. Given these considerations, and due to having exhibited robust performance during the Ego4D audio-only diarisation challenge, the Pyannote.audio-speaker-diarization-3.1\cite{Plaquet23} model was selected as the segmentation front-end for this work. However, multiple segmentation approaches were explored, and their respective results are presented in~\autoref{sec:results}.

\vspace{-0.5em} 
\subsection{Face-Voice Association with Frame-Quality-Informed Aggregation}
To attribute each speaker-invariant utterance extracted by the segmentation front-end to its corresponding visible identity, an established face-voice association framework, namely the Self-Lifting framework\cite{Self-Lifting}, was employed. This framework was selected due to its self-supervised training protocol, which is advantageous in \ac{ASD} datasets which typically lack inter-track identity annotations—i.e., identities across different tracks are often unknown, only intra-track frames are known to be identity-invariant. The Self-Lifting framework assumes self-supervision by utilising an iterative refinement process through pseudo-labelling via k-means clustering. It leverages well-known models for face and speaker recognition, specifically Inception-V1\cite{inceptionv1} and ECAPA-TDNN\cite{ecapa-tdnn}, respectively. Initially pretrained on VoxCeleb\cite{voxceleb}, the Self-Lifting framework was further finetuned in this work using the AVA-ActiveSpeaker\cite{ava-as} and Ego4D\cite{Ego4D} datasets. Self-lifting uses the multi-similarity loss function during training.

The proposed system, shown in \autoref{fig:SL_ASD}, extracts utterances from the segmentation front-end. These utterances are then input into the audio branch of the finetuned Self-Lifting model. At this point, all parameters of the Self-Lifting model are frozen. Face tracks from the corresponding video clip are then processed, with all frames fed through the visual branch of the Self-Lifting model. To enhance the robustness of the embeddings output by the visual branch, contextual information provided by each visible identities frame sequence is leveraged by a transformer encoder with learnable weights. This transformer encoder attends along the sequence dimension of the embedded faces tensor for each visible identity in the visual component of the batch. The self-attention mechanism deployed enables discriminative weighting of frame-level features through learned quality estimation, dynamically suppressing low-fidelity visual inputs while emphasising high-quality frames. Subsequently, the sequence dimension for each visible identity is aggregated via mean pooling, resulting in a single quality-informed embedding for each visible identity in the visual component of the batch.

Cross-attention is then applied, with the quality-informed embeddings serving as queries and the output of the Self-Lifting audio branch used as the keys and values. The output of the cross-attention is then fed through a single feedforward layer which collapses the embedding dimension. Here, the cross-entropy loss function is employed.

\section{Experiments}
\label{sec:experiments}
This section describes the datasets and evaluation protocol used to quantify the performance of the proposed \ac{SL-ASD} system and to compare its performance fairly with conventional synchronisation-based approaches.

\subsection{Datasets}
\label{ssec:datasets}
\textbf{Ego4D-AVD}~\cite{Ego4D} consists of recordings captured from an egocentric perspective. The dataset is composed of $572$ distinct video clips, each clip being $5$ minutes in duration, with some clips recorded concurrently. The audio was recorded by a variety of wearable devices and monaurally formatted to $16$~kHz. The video recordings were sampled at $30$~Hz in high-definition resolution. Ego4D-AVD is characterised by realistic capture conditions—including variations in illumination, occlusion, and dynamic viewpoints—which render it a challenging benchmark for \ac{ASD}. The dataset is stratified into three folds, with $379$ clips allocated for training, $50$ for validation, and $133$ for testing. Since test annotations are not provided, the training fold was further stratified into $120$ video clips for training and $23$ for development to reserve the validation fold for final evaluation. This stratification was performed to ensure that no identity appeared in more than one fold. In Ego4D, the inter-track identity annotations are consistent within video-clips. The provision of inter-track identity annotations facilitates more robust utterance-face comparisons by increasing the number of frames available for comparison, whereas the absence of such annotations would restrict comparisons to only those frames within a single track.

\subsection{Implementation Details}
\label{ssec:implementation_details}
\textbf{Front-End Utterance Segmentation:} The best performing system used in this work was the Pyannote.audio-speaker-diarization-3.1 system\cite{Plaquet23} for utterance segmentation. This segmentation was performed clipwise, and for training and development folds, it was enforced that extracted utterances had to overlap with at least $15\%$ the duration of a concurrent groundtruth utterance. This condition was not enforced during inference since groundtruth utterance boundaries would not be available in real life deployment. It was determined empirically that not imposing any limits on the minimum or maximum duration of utterances led to better performance. 

In addition to using Pyannote.audio, the Silero \ac{VAD}\cite{Silero_VAD}, in conjunction with a derivative-based speaker change detection algorithm~\cite{kartik2005speaker}, was also tested as a potential utterance segmentation front-end.

\textbf{Self-Lifting Finetuning:} During the finetuning of the Self-Lifting framework, the model was instantiated using the identical configuration to that presented in the original manuscript\cite{Self-Lifting}, except the number of cluster centroids was reduced from 1000 to 50 which better reflects the number of distinct identities in the Ego4D dataset. Only the feed forward layers after the respective pretrained encoders had parameters set to learnable during finetuning.

\textbf{\ac{SL-ASD}:} The final \ac{SL-ASD} model is configured with all parameters of the Self-Lifting framework set as frozen. The only parameters that are learnable are the transformer encoder, the cross-attention, and the feedforward layer introduced by the \ac{ASD} adaptation of Self-Lifting presented by this work. During training, the audio component of the batch comprises all utterances belonging to a single clipwise identity, while the video component of the batch is leveraged to include all face track frames for all identities within the same clip. In contrast, during validation and inference, the audio component of the batch is limited to single utterances. The Adam optimiser is used, with a learning rate of $1\times10^{-5}$ that is decayed by a factor of 0.2 every 5 epochs. A single transformer layer is utilised, and 4 attention-heads are employed for both the transformer encoder and cross-attention layers.

\subsection{Evaluation Protocol}
\label{ssec:evaluation_protocol}
Evaluation was performed using the Cartucho object detection \ac{mAP}~\cite{cartucho}, which adheres to the \ac{mAP} criterion from the PASCAL VOC2012 competition~\cite{pascal-voc-2012}. This evaluation protocol is consistent with the Ego4D audiovisual diarisation challenge~\cite{Ego4D} and recent literature~\cite{clarke23-ASD-ASRU}. Owing to the unavailability of groundtruth annotations for the test folds of Ego4D, results were reported on the validation fold, however the validation fold was not used during hyperparameter optimisation. This is in accordance with established conventions in \ac{ASD}~\cite{activespeakersincontext, ASDNet, clarke23-ASD-ASRU, SPELL, EASEE, LoCoNet}.

Due to its substantially different approach, this work necessitates a slight modification to the standard evaluation protocol. Rather than producing a framewise probability that each video frame represents an active candidate speaker, an utterance is first processed to compute the probability that each visible identity corresponds to the person who spoke the utterance. This probability is computed as a dependent measure across all visible identities in the clip. Subsequently, the face tracks that are concurrent with the utterance are identified, and the temporally corresponding video frames are assigned the computed probability of identity correspondence. 

\section{Results}
\label{sec:results}
The performance of \ac{SL-ASD} is evaluated in the following. First, the effects of different front-end segmentation systems on overall performance in terms of \ac{mAP} and utterance recall is analysed in~\autoref{tab:dynamic_eval}. Second, a comparison to synchronisation-based state-of-the-art methods in the egocentric domain is presented in~\autoref{tab:sota}.  


\begin{table}[!ht]
\centering
\caption{Performance of \ac{SL-ASD} systems and corresponding baseline evaluations on dynamic Ego4D subsets (Ego4D$\dagger$). For each \ac{SL-ASD} system, the evaluation subset is defined by the recall of the front-end utterance segmentation method, and the \ac{mAP}~[\%] is reported for the system along with TalkNet and Light-ASD baselines.}
\label{tab:dynamic_eval}
\begin{tabular}{l c c c c}
\hline
\textbf{Front-End-} & \multicolumn{3}{c}{\textbf{Ego4D$\dagger$ [\ac{mAP} \%]}} & \\ \cline{2-4}
\textbf{Method} & \ac{SL-ASD} & TalkNet~\cite{talknet} & Light-ASD~\cite{Liao_2023_CVPR}  & \multirow{-2}{*}{\textbf{Recall [\%]}} \\ \hline
Silero                 & \textbf{59.6}   & 48.7   & 58.2   & 77.5 \\
Pyannote               & \textbf{64.0}   & 55.9   & 52.8   & 91.4 \\
Groundtruth            & \textbf{80.1}   & 51.0   & 54.3   & 100  \\ \hline
\end{tabular}
\end{table}

\autoref{tab:dynamic_eval} exhibits the performance of \ac{SL-ASD}, with different front-end utterance segmentation approaches. To fully isolate the performance of \ac{SL-ASD} and illustrate the effect the utterance-recall has on the efficacy of the framework, \ac{SL-ASD} was also evaluated using groundtruth utterance boundaries. Ego4D$\dagger$ indicates a dynamic subset of the Ego4D validation fold, where `dynamic' refers to a different (likely overlapping) subset dependent on the front-end segmentation method used. Specifically, when evaluating each front-end segmentation method, only face tracks with concurrent detected utterances are used in the evaluation, all other face tracks are discarded. Effectively, \autoref{tab:dynamic_eval} illustrates two things: the recall of each front-end segmentation method, and \textemdash implicitly \textemdash the quality of the utterances extracted. False detections and speaker-variant utterances will both induce corrupted outputs from the Audio Branch of Self-Lifting, propagating errors throughout the pipeline. As results in \autoref{tab:dynamic_eval} show, \ac{SL-ASD} outperforms the baselines~\cite{talknet,Liao_2023_CVPR} on the dynamic Ego4D validation subsets (Ego4D$\dagger$).


\autoref{tab:sota} presents a comparison of the proposed \ac{SL-ASD} system with state-of-the-art \ac{ASD} methods on the Ego4D validation fold. Existing \ac{ASD} methods typically comprise several million learnable parameters, with Light-ASD's $1.0$~M parameters already being an exception, and models exceeding \ac{SL-ASD}'s performance use at least $23$~M learnable parameters. In contrast, a significantly smaller learnable parameter count of $0.40$M is achieved by the proposed \ac{SL-ASD} system.
Furthermore, to provide a more complete picture of model efficiency, the average number of floating point operations (av-FLOPs) required per inference iteration (i.e.~the number of \ac{FLOPs} per bounding box face crop in the validation set) has also been included. While the av-FLOPs for \ac{SL-ASD} remain extremely low at $0.21$~GFLOPs, comparable to Light-ASD and substantially lower than SPELL and LoCoNet, its important to note these figures are reported in isolation from shared \ac{ASD} preprocessing modules or modules reused as part of a wider implementation of a full modular \ac{AVD} pipeline~\cite{Ego4D}.

More specifically, the av-FLOPs and learnable parameters directly attributable to the face-recognition model and front-end segmenter (visual branch and segment modules in \autoref{fig:SL_ASD}, respectively) are excluded from the reported av-FLOPs and parameter count calculations. This is based on the rationale that face recognition is typically performed upstream during face track aggregation when acquiring \ac{ASD} data, and that segmentation would be performed as part of a broader modular-\ac{AVD} pipeline~\cite{Ego4D}. Consequently, the exclusion of these components is deemed justified. These results underscore that, despite its compactness, \ac{SL-ASD} is capable of achieving competitive and, in certain conditions, superior performance relative to more parameter- and compute-intensive methods. This validates the feasibility of leveraging cross-modal identity associations instead of predominantly relying on synchronisation-based approaches to \ac{ASD} in the domain of egocentric recordings.


\begin{table}[!ht]
\centering
\caption{Comparison with the state-of-the-art \ac{ASD} systems on the validation fold of Ego4D. Values for LoCoNet and SPELL are from their respective manuscripts, av-FLOPs refers to the average number of operations required to process a single bounding box face crop.}
\label{tab:sota}
\begin{tabular}{cccc}
\hline
\textbf{Method} & \textbf{mAP \%} & \textbf{\# Params. [M]} & \textbf{av-FLOPs [G]} \\ \hline
TalkNet\cite{talknet}        & 51.0                        & 16                     & 0.51 \\ 
Light-ASD\cite{Liao_2023_CVPR}       & 54.3                        & 1.0                    & \textbf{0.20} \\          
SPELL\cite{spellego4dchallenge}           & 60.7                        & 23                     &  8.7\\
LoCoNet\cite{LoCoNet}         & \textbf{68.4}               & 30                     & 0.51 \\
\hline
\ac{SL-ASD}:Silero       & 48.8                        & \textbf{0.4}                   & 0.21 \\
\ac{SL-ASD}:Pyannote     & 59.7                        & \textbf{0.4}                   & 0.21 \\
\ac{SL-ASD}:Groundtruth  & 80.1                        & \textbf{0.4}                   &  0.21 \\
\hline
\end{tabular}
\end{table}

\section{Conclusion}
\label{sec:conclusion}
This work demonstrates the feasibility of leveraging cross-modal identity associations for \ac{ASD} in egocentric recordings. It has been shown that, by integrating an adapted face-voice association model with a robust utterance segmentation front-end and a transformer-based temporal aggregator, the proposed \ac{SL-ASD} system is capable of achieving performance that is competitive with, and in certain conditions superior to, much more parameter-intensive synchronisation-based methods. A significant reduction in the number of learnable parameters has been attained, when excluding non-learnable components that are presumed to be inherently present in modular audiovisual diarisation pipelines. Overall, it is concluded that the reliance on temporal synchronisation can be mitigated by exploiting biometric associations, which proves to be a viable approach for addressing the challenges posed by occlusions, motion blur, and noisy acoustic conditions in egocentric scenarios. 

\bibliographystyle{IEEEtran}
\bibliography{mybib}

\begin{thebibliography}{10}
\providecommand{\url}[1]{#1}
\csname url@samestyle\endcsname
\providecommand{\newblock}{\relax}
\providecommand{\bibinfo}[2]{#2}
\providecommand{\BIBentrySTDinterwordspacing}{\spaceskip=0pt\relax}
\providecommand{\BIBentryALTinterwordstretchfactor}{4}
\providecommand{\BIBentryALTinterwordspacing}{\spaceskip=\fontdimen2\font plus
\BIBentryALTinterwordstretchfactor\fontdimen3\font minus \fontdimen4\font\relax}
\providecommand{\BIBforeignlanguage}[2]{{%
\expandafter\ifx\csname l@#1\endcsname\relax
\typeout{** WARNING: IEEEtran.bst: No hyphenation pattern has been}%
\typeout{** loaded for the language `#1'. Using the pattern for}%
\typeout{** the default language instead.}%
\else
\language=\csname l@#1\endcsname
\fi
#2}}
\providecommand{\BIBdecl}{\relax}
\BIBdecl

\bibitem{activespeakersincontext}
J.~L. Alcazar, F.~C. Heilbron, L.~Mai, F.~Perazzi, J.-Y. Lee, P.~Arbel{\'a}ez, and B.~Ghanem, ``{Active Speakers in Context},'' \emph{2020 IEEE/CVF Conference on Computer Vision and Pattern Recognition (CVPR)}, 2020.

\bibitem{ava-as}
J.~Roth, S.~Chaudhuri, O.~Klejch, R.~Marvin, A.~Gallagher, L.~Kaver, S.~Ramaswamy, A.~Stopczynski, C.~Schmid, Z.~Xi, and C.~Pantofaru, ``{Ava Active Speaker: An Audio-Visual Dataset for Active Speaker Detection},'' in \emph{Proc.~Int.~Conf.~on Acoustics, Speech and Signal Processing (ICASSP)}, 2020.

\bibitem{asdtransformer}
G.~Datta, T.~Etchart, V.~Yadav, V.~Hedau, P.~Natarajan, and S.-F. Chang, ``{ASD-Transformer: Efficient Active Speaker Detection Using Self And Multimodal Transformers},'' in \emph{Proc.~IEEE Int.~Conf.~on Acoustics, Speech and Signal Processing (ICASSP)}, 2022.

\bibitem{buffy}
M.~Everingham, J.~Sivic, and A.~Zisserman, ``{Hello! My name is... Buffy'' -- Automatic Naming of Characters in TV Video},'' in \emph{British Machine Vision Conference}, 2006.

\bibitem{talknet}
R.~Tao, Z.~Pan, R.~K. Das, X.~Qian, M.~Z. Shou, and H.~Li, ``{Is Someone Speaking? Exploring Long-term Temporal Features for Audio-visual Active Speaker Detection},'' in \emph{Proc.~29th ACM Int.~Conf.~on Multimedia}, 2021.

\bibitem{Liao_2023_CVPR}
J.~Liao, H.~Duan, K.~Feng, W.~Zhao, Y.~Yang, and L.~Chen, ``{A Light Weight Model for Active Speaker Detection},'' in \emph{Proc.~IEEE/CVF Conf.~on Computer Vision and Pattern Recognition (CVPR)}, June 2023.

\bibitem{ASDNet}
O.~Köpüklü, M.~Taseska, and G.~Rigoll, ``{How to Design a Three-Stage Architecture for Audio-Visual Active Speaker Detection in the Wild},'' in \emph{2021 IEEE/CVF International Conference on Computer Vision (ICCV)}, 2021.

\bibitem{sync-talknet}
A.~Wuerkaixi, Y.~Zhang, Z.~Duan, and C.~Zhang, ``Rethinking audio-visual synchronization for active speaker detection,'' in \emph{Proc.~32nd IEEE Int.~Conf.~on Machine Learning for Signal Processing}, 2022.

\bibitem{lookwhostalking}
R.~Cutler and L.~S. Davis, ``{Look who's talking: speaker detection using video and audio correlation},'' \emph{2000 IEEE International Conference on Multimedia and Expo. ICME2000. Proceedings. Latest Advances in the Fast Changing World of Multimedia (Cat. No.00TH8532)}, vol.~3, 2000.

\bibitem{EASEE}
J.~L. Alcazar, M.~Cordes, C.~Zhao, and B.~Ghanem, ``{End-to-End Active Speaker Detection},'' in \emph{European Conference on Computer Vision}, 2022.

\bibitem{SPELL}
K.~Min, S.~Roy, S.~Tripathi, T.~Guha, and S.~Majumdar, ``{Learning Long-Term Spatial-Temporal Graphs for Active Speaker Detection},'' in \emph{Euro.~Conf.~on Computer Vision}, 2022.

\bibitem{LeonAlcazar2021MAASMA}
J.~Le'on-Alc'azar, F.~C. Heilbron, A.~K. Thabet, and B.~Ghanem, ``{MAAS: Multi-modal Assignation for Active Speaker Detection},'' \emph{2021 IEEE/CVF International Conference on Computer Vision (ICCV)}, 2021.

\bibitem{clarke23-ASD-ASRU}
J.~Clarke, Y.~Gotoh, and S.~Goetze, ``{Improving Audiovisual Active Speaker Detection in Egocentric Recordings with the Data-Efficient Image Transformer},'' in \emph{IEEE Automatic Speech Recognition and Understanding Workshop (ASRU23)}, 2023.

\bibitem{TS-talknet}
Y.~Jiang, R.~Tao, Z.~Pan, and H.~Li, ``{Target Active Speaker Detection with Audio-visual Cues},'' in \emph{Proc. Interspeech}, 2023.

\bibitem{clarke2025speakerembeddinginformedaudiovisual}
J.~Clarke, Y.~Gotoh, and S.~Goetze, ``{Speaker Embedding Informed Audiovisual Active Speaker Detection for Egocentric Recordings},'' in \emph{Int.\ Conf. on Acoustics, Speech and Signal Processing (ICASSP)}, 2025.

\bibitem{Ego4D}
``{Ego4D: Around the World in 3,000 Hours of Egocentric Video},'' \emph{2022 IEEE/CVF Conference on Computer Vision and Pattern Recognition (CVPR)}, 2021.

\bibitem{spellego4dchallenge}
\BIBentryALTinterwordspacing
T.~Ishibashi, K.~Ono, N.~Kugo, and Y.~Sato, ``{Technical Report for Ego4D Long Term Action Anticipation Challenge 2023},'' 2023. [Online]. Available: \url{https://arxiv.org/abs/2307.01467}
\BIBentrySTDinterwordspacing

\bibitem{seeking_shape_of_sound}
P.~Wen, Q.~Xu, Y.~Jiang, Z.~Yang, Y.~He, and Q.~Huang, ``Seeking the shape of sound: An adaptive framework for learning voice-face association,'' in \emph{Proceedings of the IEEE/CVF Conference on Computer Vision and Pattern Recognition (CVPR)}, June 2021, pp. 16\,347--16\,356.

\bibitem{Self-Lifting}
\BIBentryALTinterwordspacing
G.~Chen, D.~Zhang, T.~Liu, and X.~Du, ``Self-lifting: A novel framework for unsupervised voice-face association learning,'' in \emph{Proceedings of the 2022 International Conference on Multimedia Retrieval}, ser. ICMR '22.\hskip 1em plus 0.5em minus 0.4em\relax New York, NY, USA: Association for Computing Machinery, 2022, p. 527–535. [Online]. Available: \url{https://doi.org/10.1145/3512527.3531364}
\BIBentrySTDinterwordspacing

\bibitem{learnable_pins}
\BIBentryALTinterwordspacing
A.~Nagrani, S.~Albanie, and A.~Zisserman, ``Learnable pins: Cross-modal embeddings for person identity,'' in \emph{Computer Vision – ECCV 2018: 15th European Conference, Munich, Germany, September 8-14, 2018, Proceedings, Part XIII}.\hskip 1em plus 0.5em minus 0.4em\relax Berlin, Heidelberg: Springer-Verlag, 2018, p. 73–89. [Online]. Available: \url{https://doi.org/10.1007/978-3-030-01261-8_5}
\BIBentrySTDinterwordspacing

\bibitem{single-branch}
M.~S. Saeed, S.~Nawaz, Yousaf, M.~H. Khan, M.~Z. Zaheer, K.~Nandakumar, M.~H. Yousaf, and A.~Mahmood, ``Single-branch network for multimodal training,'' in \emph{ICASSP 2023-2023 IEEE International Conference on Acoustics, Speech and Signal Processing (ICASSP)}.\hskip 1em plus 0.5em minus 0.4em\relax IEEE, 2023.

\bibitem{disentangled_cross-modal_biometric_matching}
H.~Ning, X.~Zheng, X.~Lu, and Y.~Yuan, ``Disentangled representation learning for cross-modal biometric matching,'' \emph{IEEE Transactions on Multimedia}, vol.~24, pp. 1763--1774, 2022.

\bibitem{momentum_face-voice}
Y.~Qiu, Z.~Yu, and Z.~Gao, ``An efficient momentum framework for face-voice association learning,'' in \emph{Pattern Recognition and Computer Vision}, Q.~Liu, H.~Wang, Z.~Ma, W.~Zheng, H.~Zha, X.~Chen, L.~Wang, and R.~Ji, Eds.\hskip 1em plus 0.5em minus 0.4em\relax Singapore: Springer Nature Singapore, 2024, pp. 271--283.

\bibitem{unimodal_similarity}
R.~Sharma and S.~Narayanan, ``Audio-visual activity guided cross-modal identity association for active speaker detection,'' \emph{IEEE Open Journal of Signal Processing}, vol.~4, pp. 225--232, 2023.

\bibitem{FaVoA}
\BIBentryALTinterwordspacing
H.~Carneiro, C.~Weber, and S.~Wermter, ``Favoa: Face-voice association favours ambiguous speaker detection,'' in \emph{Artificial Neural Networks and Machine Learning – ICANN 2021: 30th International Conference on Artificial Neural Networks, Bratislava, Slovakia, September 14–17, 2021, Proceedings, Part I}.\hskip 1em plus 0.5em minus 0.4em\relax Berlin, Heidelberg: Springer-Verlag, 2021, p. 439–450. [Online]. Available: \url{https://doi.org/10.1007/978-3-030-86362-3_36}
\BIBentrySTDinterwordspacing

\bibitem{what_do_we_hear}
W.~Gaver, ``What in the world do we hear?: An ecological approach to auditory event perception,'' \emph{Ecological Psychology}, vol.~5, pp. 1--29, 03 1993.

\bibitem{infant_fva}
L.~Bahrick, M.~Hernandez-Reif, and R.~Flom, ``The development of infant learning about specific face–voice relations,'' \emph{Developmental psychology}, vol.~41, pp. 541--52, 05 2005.

\bibitem{Plaquet23}
A.~Plaquet and H.~Bredin, ``{Powerset multi-class cross entropy loss for neural speaker diarization},'' in \emph{Proc. INTERSPEECH 2023}, 2023.

\bibitem{inceptionv1}
\BIBentryALTinterwordspacing
C.~Szegedy, W.~Liu, Y.~Jia, P.~Sermanet, S.~E. Reed, D.~Anguelov, D.~Erhan, V.~Vanhoucke, and A.~Rabinovich, ``Going deeper with convolutions,'' \emph{2015 IEEE Conference on Computer Vision and Pattern Recognition (CVPR)}, pp. 1--9, 2014. [Online]. Available: \url{https://api.semanticscholar.org/CorpusID:206592484}
\BIBentrySTDinterwordspacing

\bibitem{ecapa-tdnn}
\BIBentryALTinterwordspacing
B.~Desplanques, J.~Thienpondt, and K.~Demuynck, ``{ECAPA-TDNN: Emphasized Channel Attention, Propagation and Aggregation in TDNN Based Speaker Verification},'' in \emph{Interspeech 2020}.\hskip 1em plus 0.5em minus 0.4em\relax ISCA, Oct 2020. [Online]. Available: \url{http://dx.doi.org/10.21437/Interspeech.2020-2650}
\BIBentrySTDinterwordspacing

\bibitem{voxceleb}
\BIBentryALTinterwordspacing
A.~Nagrani, J.~S. Chung, and A.~Zisserman, ``{VoxCeleb: A Large-Scale Speaker Identification Dataset},'' in \emph{Interspeech 2017}.\hskip 1em plus 0.5em minus 0.4em\relax ISCA, Aug. 2017. [Online]. Available: \url{http://dx.doi.org/10.21437/Interspeech.2017-950}
\BIBentrySTDinterwordspacing

\bibitem{Silero_VAD}
S.~Team, ``{Silero VAD: pre-trained enterprise-grade Voice Activity Detector (VAD), Number Detector and Language Classifier},'' \url{https://github.com/snakers4/silero-vad}, 2024.

\bibitem{kartik2005speaker}
V.~Kartik, D.~S. Satish, and C.~C. Sekhar, ``Speaker change detection using support vector machines,'' in \emph{Proceedings of NOLISP}, 2005, pp. 22--25.

\bibitem{cartucho}
J.~{Cartucho}, R.~{Ventura}, and M.~{Veloso}, ``{Robust Object Recognition Through Symbiotic Deep Learning In Mobile Robots},'' in \emph{2018 IEEE/RSJ International Conference on Intelligent Robots and Systems (IROS)}, 2018.

\bibitem{pascal-voc-2012}
M.~Everingham, L.~Van~Gool, C.~Williams, J.~Winn, and A.~Zisserman, ``The {PASCAL} {V}isual {O}bject {C}lasses {C}hallenge 2012 {(VOC2012)} {R}esults,'' \url{http://www.pascal-network.org/challenges/VOC/voc2012/workshop/index.html}.

\bibitem{LoCoNet}
X.~Wang, F.~Cheng, and G.~Bertasius, ``{LoCoNet: Long-Short Context Network for Active Speaker Detection},'' in \emph{Proceedings of the IEEE/CVF Conference on Computer Vision and Pattern Recognition (CVPR)}, June 2024.

\end{thebibliography}

\end{document}